\RequirePackage{fix-cm}
\documentclass[smallextended]{svjour3}       
\smartqed  
\usepackage{graphicx}
\usepackage{hyperref,amsmath}

\newcommand{\be}{\begin{equation}}
\newcommand{\ee}{\end{equation}}
\newcommand{\I}{{\cal I}}
\newcommand{\da}{\downarrow}
\newcommand{\ua}{\uparrow}
 \journalname{
Journal of Statistical Physics
}
\begin{document}

\title{A remark on the notion of independence \\
of quantum integrals of motion\\ in the thermodynamic limit
}


\author{Oleg Lychkovskiy
}


\institute{   Skolkovo Institute of Science and Technology, Bolshoy Boulevard 30, bld. 1
 Moscow 121205, Russia
 \\
Steklov Mathematical Institute of Russian Academy of Sciences,  Gubkina str. 8, Moscow 119991, Russia\\
              \email{o.lychkovskiy@skoltech.ru}           
}

\date{Received: date / Accepted: date}

\maketitle

\begin{abstract}
Studies of integrable quantum many-body systems have a long history with an impressive record of success. However, surprisingly enough, an unambiguous definition of quantum integrability remains a matter of an ongoing debate.  We contribute to this debate by dwelling upon an important aspect of quantum integrability -- the notion of independence of quantum integrals of motion (QIMs). We point out that a widely accepted definition of functional independence of QIMs is flawed, and suggest a new definition. Our study is motivated by  the PXP model -- a model of $N$ spins $1/2$ possessing an extensive number of binary QIMs. The number of QIMs which are independent according to the common definition turns out to be equal to the number of spins, $N$. A common wisdom would then suggest that the system is completely integrable, which is not the case. We discuss the origin of this conundrum and demonstrate how it is resolved when a new definition of independence of QIMs is employed.
\end{abstract}

\section{Introduction\label{sec: introduction}}

A classical Hamiltonian system with $N$ degrees of freedom is said to be completely integrable if it posses $N$ functionally independent integrals of motion in involution (i.e. with pairwise commuting Poisson brackets). This concise, clear and rigorous definition is a mainstay of the well-developed and very fruitful theory of classical integrability. The situation with quantum integrability is remarkably different: It is fair to say that even a commonly accepted rigorous definition of quantum integrability is lacking. An attempt of a straightforward translation of the classical definition to the quantum language stumbles upon several ambiguities: How to define and count quantum ``degrees of freedom'' \cite{zhang1989,weigert1992,faddeev2007}? What notion of independence of quantum integrals of motion (QIMs) should be used \cite{Sklyanin1992,gravel2002,Tempesta2004}? Should proper integrals of motion be local in some sense \cite{caux2011remarks}? While a number of working definitions of quantum integrability are in use \cite{Weigert1995,gravel2002,caux2011remarks,sutherland2004beautiful,owusu2011,owusu2013}, none of them is of the same level of rigor, generality and usefulness as the definition of the classical integrability. The very existence of several different definitions indicates that the issue is not settled.

It is expected that integrable and non-integrable quantum many-body systems are markedly different in a number of  aspects. The list of differences includes the (non)existence of an extensive number of local integrals of motion, level statistics (Wigner-Dyson \cite{bohigas} {\it vs} Poisson \cite{berry}), (non)validity of the eigenstate thermalization hypothesis \cite{deutsch,srednicki,rigol,kim2014testing} and the canonical universality hypothesis \cite{dymarsky}, the nature of the local steady state approached after relaxation from a non-equilibrium initial state (Gibbs thermal state {\it vs}  generalized Gibbs state) \cite{vidmar2016generalized,mori2018thermalization}. These expectations are rooted in a huge amount of analytical and numerical work pertaining to specific systems as well as in insights from the random matrix theory \cite{mori2018thermalization}.  This body of work shows that typically a system which possesses one of the properties in the list, also possesses others. On the integrable side, this is verified for quadratic fermionic and bosonic systems as well as for systems solvable by Bethe ansatz. On the nonintegrable side, these properties have been tested in many numerical simulations.  The experience gained in these studies allows one to informally classify almost any specific system as integrable or nonintegrable. Still, a general proof of the above-mentioned attributes of (non)integrability based on some formal definition is lacking.

Here we attempt to clarify one of the issues of the formal definition of integrability mentioned above. Namely,  we focus on the notion of independence of QIMs.  Our work is inspired by a particular model of $N$ spins 1/2 known as the PXP model \cite{lesanovsky2012interacting,bernien2017probing,turner2018weak}. This model posses an extensive number of local QIMs. We first attempt to identify a set of QIMs which are mutually independent according to the widely accepted definition. The latter definition essentially classifies a set of QIMs as functionally independent unless one of the integrals can be expressed as a function of the others. According to this definition, the PXP model turns out to posses $N$ functionally independent local binary\footnote{i.e. having only two different eigenvalues} integrals of motion. One thus might expect that the PXP model is completely integrable and, in particular, all $2^N$ eigenstates can be unambiguously enumerated by  $2^N$ different combinations of the eigenvalues of its $N$ binary integrals of motion.  However, this is not the case: The PXP model is not integrable (in particular, it has a Wigner-Dyson level statistics in a large subspace of the Hilbert space \cite{turner2018weak}), and the above-mentioned enumeration can not be carried out. Furthermore, certain polynomials constructed from these ``independent'' QIMs turn out to be identically zero, which is at odds with the intuitive notion of independence.

This conundrum motivates us to reconsider the notion of independence of QIMs. We suggest a new definition of this notion which allows for a more general functional dependence of integrals of motion, $I_1,I_2,...,I_M$, of the form $F(I_1,I_2,...,I_M)=0$. The triviality of this generalization is deceptive: It would not work unless supplemented by certain requirements imposed on the function~$F$. To be specific, one must exclude a class of functions~$F$ which is trivial in a certain sense, and, further, require that the complexity of $F$ is polynomial in the system size. We incorporate these requirements in our definition.

Armed with the new definition of independence of QIMs, we then return to the PXP model. We prove that according to this new definition there are still (at least) $2N/3$ commuting local independent binary QIMs.  Thus the PXP model constitutes an example of  a nonintegrable quantum many-body model with an extensive number of independent local integrals of motion in involution.

The rest of the paper is organised as follows. In the next section we introduce the PXP model. In Sect. \ref{sec: common} we review the common definition of the independence of QIMs, demonstrate its failure in the PXP model, and expose the reasons for this failure. In Sect. \ref{sec: new} we give a new definition of the independence of QIMs and demonstrate how it evades various pitfalls discussed in the present paper and elsewhere. In particular, we identify a subset of  QIMs of the PXP model which are independent according to this definition. Sect. \ref{sec: summary} contains summary and concluding remarks.

\section{PXP model and its integrals of motion}
We consider a one-dimensional chain of $N$ spins $1/2$ with a translation-invariant Hamiltonian
\be\label{H}
H=\sum_{i=1}^{N} P_i\, \sigma_{i+1}^x \, P_{i+2},
\ee
where the projection operator $P_i$ reads
\be\label{P}
P_i\equiv\frac12 (1-\sigma_i^z),
\ee
and lattice sites are enumerated modulo $N$.
This model (or its close proxies) can describe bosons on a lattice with next-nearest hard-core constraints  \cite{sachdev2002mott,fendley2004competing} or Rydberg atoms on a lattice subject to dipole blockade preventing simultaneous excitation of two neighboring atoms \cite{lesanovsky2012interacting,bernien2017probing}. Furthermore, a one-dimensional version of the two-fermion model of high-temperature superconductivity in cuprates \cite{fine2004hypothesis,fine2005temperature,fine2007magnetic,bhartiya2017superconductivity} can be mapped to the Hamiltonian \eqref{H} in a certain limit.


The three-body interaction in eq. \eqref{H} flips a spin provided that two neighboring spins are down (this way it is able to describe  hard core or Rydberg blockade constraints). Therefore, if two neighboring spins happen to be up, they will stay up forever.   As a consequence, $N$ projection operators
\be\label{I}
I_i \equiv (1-P_i)(1-P_{i+1})=\frac{1+\sigma_i^z}{2}\,\frac{1+\sigma_{i+1}^z}{2},~~~ i=1,2,..., N,
\ee
satisfy
\be\label{IH}
HI_i=I_i H=[H,I_i]=0
\ee
and are thus integrals of motion. Furthermore, these $N$ integrals of motion are in involution, i.e.
\be\label{involution}
[I_i,I_j]=0~~~~\forall i,j.
\ee

Despite the existence of the above integrals of motion, the model \eqref{H} should be classified as nonintegrable \cite{turner2018weak} for the following reasons. Consider first the sector of the Hilbert space with all $I_i$ equal to zero (no neighboring spins pointing up).\footnote{
When the Hamiltonian \eqref{H} is applied to Rydberg atoms or hard-core bosons, this is the only physical sector \cite{lesanovsky2012interacting,bernien2017probing,sachdev2002mott,fendley2004competing}.
In our studies we ignore this physical context and consider the Hamiltonian \eqref{H} as acting in the full Hilbert space of dimension $2^N$.}
This sector has an exponentially large Hilbert space \cite{lesanovsky2012interacting}.  In this sector the level statistics of the model exhibits level repulsion which is a clear footprint of nonintegrability \cite{turner2018weak}.  The same is expected when a single $I_i$ is equal to one and others are zero. When two non-neighboring $I_i$ are equal to one, the chain is effectively cut in two noninteracting pieces by two stable $\uparrow\uparrow$ configurations along the chain. In other words, the system consists of two noninteracting nonintegrable systems. It is natural to count this case as nonintegrable, too. The same logic holds as long as the number of cuts (i.e. $I_i$ equal to 1) is $o(N)$. Thus  the system can be firmly  classified as nonintegrable in the large portion of the Hilbert space. An additional argument in favour of nonintegrability is that the system thermalizes, albeit the thermalization time scale can be atypically large \cite{bernien2017probing,turner2018weak,turner2018quantum}.

Given the existence of $N$ binary QIMs in involution \eqref{I} on the one hand and the nonintegrability of the model on the other, it is natural to ask how many {\it independent} integrals of motion the PXP model possesses.  This question is addressed in the next two sections.


\section{Common definition of independence of QIMs \label{sec: common}}

In the present section and in Sect. \ref{sec: new} we dwell upon the notion of independence of quantum integrals of motion. Our discussion is motivated by but not restricted to the specific model \eqref{H}. Whenever our discussion is general, the notations $H$ and $I_1,I_2,...,I_M$ are unrelated to the specific Hamiltonian \eqref{H} and QIMs \eqref{I} but refer to a general many-body quantum Hamiltonian and its $M$  integrals of motion, respectively. It should be stressed, however, that we consider only sets of QIMs in involution, i.e. eq. \eqref{involution} is always valid. We use a notation $\I_i$ for an eigenvalue of $I_i$.
Whether the notations $H$, $I_i$ and $\I_i$ apply to the specific model \eqref{H} or to a general case  will be clear from the context.

We start from reviewing a commonly used \cite{gravel2002,caux2011remarks,ros2015integrals} notion of functional independence of QIMs which we refer to as {\it independence in the weak sense} or, briefly, weak independence.

\medskip

\noindent {\it Definition 1.} Commuting integrals of motion $I_1,I_2,...,I_M$ are functionally independent in the weak sense unless  one of the integrals, $I_i$, can be expressed as a function of others.

\medskip

Note that since all $I_i$ commute and thus share a common eigenbasis, the notion of function of operators here is free from the ordering ambiguities. Sometimes a class of functions $F$ is restricted to polynomials in which case thus defined independence is referred to as algebraic \cite{gravel2002,caux2011remarks}.

It is easy to see that all $N$ QIMs \eqref{I} of the model \eqref{H} are independent in the weak sense. Indeed, assume the opposite, e.g. that
$I_1=F(I_2,I_3,...,I_N)$ for some function $F$. This implies that for any two common eigenstates of QIMs, $\Psi$ and $\Psi'$, with $(N-1)$  coinciding eigenvalues $\I_2,\I_3,...,\I_N$ of the integrals $I_2,I_3,...,I_N$, the respective eigenvalues of $I_1$ also coincide and equal $F(\I_2^{\mu_2},\I_3^{\mu_3},...,\I_N^{\mu_N})$. This is not the case, as can be readily verified e.g. for  $N=3$, $\Psi=|\uparrow\uparrow\downarrow\rangle$ and $\Psi'=|\downarrow\uparrow\downarrow\rangle$, see Table \ref{table}  (the generalization to higher $N$ is straightforward).

\begin{table}
\caption{Eigenvalues $\{\I_1,\I_2,\I_3\}$ and eigenstates of integrals of motion \eqref{I} of the PXP model \eqref{H} with $N=3$. One combination of eigenvalues, $\I_1= \I_2=\I_3=0$,  corresponds to four eigenstates, while three combinations of eigenvalues with $\I_1+\I_2+\I_3=2$  are forbidden by the relation \eqref{dependences} and thus do not correspond to any eigenstate. Note that the eigenstates of the Hamiltonian in the sector with $\I_1= \I_2=\I_3=0$ are linear combinations of the states shown in the table. }
\label{table}       
\begin{tabular}{l|llllllll}
\hline\noalign{\smallskip}
$\{\I_1,\I_2,\I_3\}$ & $\{0,0,0\}$  & $\{1,0,0\}$ & $\{0,1,0\}$ & $\{0,0,1\}$ & $\{1,1,0\}$ & $\{1,0,1\}$ & $\{0,1,1\}$ & $\{1,1,1\}$ \\
\noalign{\smallskip}\hline\noalign{\smallskip}
eigenstates & $|\da\da\da\rangle$ & $|\ua\ua\da\rangle$ & $|\da\ua\ua\rangle$ & $|\ua\da\ua\rangle$ & --- & --- & --- & $|\ua\ua\ua\rangle$  \\
        & $|\ua\da\da\rangle$ &  & & & & & & \\
        & $|\da\ua\da\rangle$ &  & & & & & & \\
        & $|\da\da\ua\rangle$ &  & & & & & & \\
\noalign{\smallskip}\hline
\end{tabular}
\end{table}

When a system of $N$ spins $1/2$ possesses $N$ local binary QIMs classified as independent, one might expect that this model is completely integrable according to any reasonable notion of complete integrability. This expectation is based in a tacit assumption that all $2^N$ eigenstates can be unambiguously enumerated by  $2^N$ different combinations of the eigenvalues of $N$ integrals of motion. This would imply that the Hilbert space is partitioned in a tensor product of $N$ two-dimensional Hilbert spaces, and the quantum many-body dynamics reduces to the dynamics of $N$ decoupled two-level systems. This indeed happens e.g. in systems described by quadratic  fermionic Hamiltonians (in particular, in the integrable $XY$ model of spins $1/2$ \cite{lieb1961two}).\footnote{It should be stressed that the notion of the number of independent QIMs is not meaningful  without referring to the number of eigenvalues of each QIM. For example, one could substitute every couple of binary QIMs $I_{2i}$, $I_{2i+1}$ by a quaternary QIM
$(I_{2i}+2 I_{2i+1})$
with the eigenvalues $\{0,1,2,3\}$, and thus reduce the number of charges twofold.  }

Yet, the model \eqref{H} is undeniably {\it not} completely integrable, as discussed in Sect. \ref{sec: introduction}, and the above mentioned enumeration can not be carried out, see Table \ref{table}. The catch is that the independence in the weak sense as determined by {\it Definition 1} fails to grasp the intuitive meaning of independence of QIMs. In fact, there are $N$ relations between the QIMs \eqref{I} of the form
\be\label{dependences}
I_i(I_{i+1}-1)I_{i+2}=0, ~~~~~~~~i=1,2,...,N.
\ee
There is no way to express one QIM as a function of the others using these relations, thus they do not invalidate the independence of QIMs \eqref{I} in the weak sense. Still, they restrict possible combinations of eigenvalues of QIMs so that the total number of allowed combinations is less than $2^N$. This is to say, certain combinations of eigenvalues correspond to several different eigenstates  each (these eigenstates form invariant subspaces), while other combinations do not correspond to any eigenstate, see Table \ref{table}. This is the reason why the eigenvalues of $N$ QIMs \eqref{I} fall short in enumerating a complete basis in the full Hilbert space of the PXP model. Furthermore, the dimensions of the invariant subspaces in the combined spectrum of the set of QIMs \eqref{I} in general grow exponentially with $N$, thus leaving enough room for the full-fledged nonintegrability to develop.

It should be noted, however, that despite the above-discussed deficiency of the notion the independence of QIMs in the weak sense, such weak independence remains a meaningful characteristics of a set of QIMs. Indeed, consider again the above simple example of the PXP model with  $N=3$ spins. Three weakly independent QIMs foliate the Hilbert space in the five invariant subspaces, see Table \ref{table}. However, if we discard one of three QIMs, we will loose some information on the eigenbasis of the model, since the remaining two QIMs will foliate the Hilbert space in only four subspaces. This illustrates that all QIMs independent in the weak sense should be taken into account to attain the maximal foliation of the Hilbert space in invariant subspaces, even if they are not independent in the strong sense defined in what follows.


\section{New definition of independence of QIMs\label{sec: new}}

\subsection{Naive generalisation of the notion of independence of QIMs}

The above discussion exposes the need for a more elaborated definition of independence of quantum integrals of motion. Motivated by the relation \eqref{dependences}, one might be tempted to introduce the following

\medskip

\noindent {\it Naive definition.} Integrals of motion $I_1,I_2,...,I_M$ are functionally independent unless there exists a function $F$ such that $F(I_1,I_2,...,I_M)=0$.

\medskip

While the need to account for the functional dependence of a general form is well understood \cite{gravel2002}, the application of this naive definition is hindered by two quite well-known pitfalls. In the next two subsections we discuss them and demonstrate how they can be remedied. These remedies, being incorporated into the naive definition, turn the latter into a valid one. This new definition is stated in  Sect. \ref{subsec}.

\subsection{Triviality of functional dependence}

The first pitfall is that there exist ``trivial'' functions $F$ such that while the relation $F(I_1,I_2,...,I_M)=0$ holds, it does not imply any actual dependence of QIMs. For example, the very fact that $I_i$ defined by eq. \eqref{I} are projection operators leads to a trivial relation
\be\label{very trivial dependence}
I_i^2-I_i=0,~~~~~~~~i=1,2,...,N,
\ee
and to a plethora of apparently more involved but still trivial relations, e.g.
\be\label{trivial dependence}
I_iI_{i+1}(I_{i+1}-I_iI_{i+2})I_{i+2}=0, ~~~~~~~~i=1,2,...,N.
\ee
Clearly, such relations are irrelevant for  the task of determining the number of independent QIMs.

It is important to realize that there is a crucial difference between the functional dependence \eqref{dependences} on the one hand and \eqref{very trivial dependence}, \eqref{trivial dependence} on the other: The relations \eqref{very trivial dependence}, \eqref{trivial dependence} hold when we substitute the integrals of motion by an {\it arbitrary} combination of the corresponding eigenvalues, while the relation \eqref{dependences}  does not. Indeed, eqs. \eqref{very trivial dependence}, \eqref{trivial dependence} are mere consequences of the fact that
\begin{align}
x^2-x=0~~{\rm and}~~x y(y-xz)z=0 ~~~~~\forall~ x,y,z\in\{0,1\}.
\end{align}
In contrast, there exists a combination $\{\I_i=1,\I_{i+1}=0,\I_{i+2}=1\}$ of eigenvalues of operators $I_i,I_{i+1},I_{i+2}$ which does not satisfy the functional relation \eqref{dependences}, i.e.
\be
\I_i(\I_{i+1}-1)\I_{i+2}\neq0.
\ee
This means that eq. \eqref{dependences} imposes  a restriction on compatibility of eigenvalues ({\it cf.} Table \ref{table} and discussion in Sect. \ref{sec: common}) and thus should be regarded as a nontrivial functional dependence. These arguments motivate us to introduce

\medskip

\noindent {\it Definition 2.} The functional dependence $F(I_i,I_i,...,I_M)=0$ of commuting integrals of motion $I_i,I_i,...,I_M$ is {\it trivial} if $F(\I_i,\I_i,...,\I_M)=0$ for an arbitrary combination $\I_i,\I_i,...,\I_M$ of eigenvalues of $I_i,I_i,...,I_M$, and {\it nontrivial} otherwise.
\medskip

\noindent We reiterate that according to this definition eq. \eqref{dependences} determines a nontrivial functional dependence between QIMs \eqref{I}, while eqs. \eqref{very trivial dependence} and \eqref{trivial dependence} -- trivial ones.

\smallskip

It should be stressed that the notion of triviality introduced by the {\it Definition 2} refers not to the function $F$ alone, but collectively to the function $F$ and the set of QIMs $I_i,$ $i=1,2,\dots,M$.

\subsection{Complexity of  functional dependence}
To expose the second pitfall of the naive definition of the independence of QIMs, we review an issue first raised in \cite{weigert1992}. This issue is based on a theorem by von Neumann \cite{neumann1931uber} which states that  any number of commuting operators  can be expressed as functions of a certain other operator. In particular, if one of the integrals of motion, say, $I_1$, has a non-degenerate eigenvalue spectrum (this is not uncommon for integrable Hamiltonians), then any other integral of motion $I_i$ which commutes with $I_1$ can be expressed as a polynomial of $I_1$ as follows:
\be\label{Neumann 1}
I_i=\sum_{\mu=1}^d \I_i^\mu\, \frac{\prod_{\nu\neq\mu} (I_1-\I_1^\nu)}{\prod_{\nu\neq\mu} (\I_1^\mu-\I_1^\nu)}.
\ee
Here $d$ is the dimension of the Hilbert space, $\I_1^\mu$ and $\I_i^\mu$ are eigenvalues of respectively $I_1$ and $I_i$ in the common eigenbasis $|\mu\rangle,$ $\mu=1,2,\dots,d$, and  index $\nu$ in the products runs over $d-1$ integers distinct from $\mu$.

If we were to accept the functional dependence of QIMs given by eq. \eqref{Neumann 1} as relevant, we would be left with a single independent integral of motion, $I_1$, and thus never get an integrable system. This is the essence of the issue discussed in \cite{weigert1992}. Note that the assumption of a nondegenerate spectrum of $I_1$ is not essential for the argument and can be avoided \cite{weigert1992}.\footnote{For example, given a set $I_1,I_2,\dots,I_M$ of $M$ mutually commuting QIMs, one can replace $I_1$ by a new QIM  $I_{\Sigma}$  given by
\be
I_{\Sigma}=\sum_{i=1}^M c_i I_i.
\ee
One can always choose numerical coefficients $c_i$ in such a way that $I_{\Sigma}$ has a nondegenerate spectrum.}  Note also that
the functional dependence \eqref{Neumann 1} is nontrivial according to the {\it Definition~2}.\footnote{The awkwardness of the functional dependence  \eqref{Neumann 1} is particularly stunning for a translation-invariant system of fermions with periodic boundary conditions. The Hamiltonian and the total momentum are two integrals of motion of this system, which are generally regarded as independent. However, if the accidental degeneracies of the Hamiltonian are absent (this is generically the case for interacting fermions as well as for noninteracting fermions in the presence of a magnetic flux), the operator of the total momentum can be expressed as a function of the Hamiltonian.     }

The reason while the  functional dependence \eqref{Neumann 1} is irrelevant for the definition of integrability, at least in the many-body context, is its exponential complexity. This is to say, the right hand side of eq. \eqref{Neumann 1} is a polynomial of power $(d-1)$  with $(d-1)$ nontrivial coefficients, and the dimension $d$ of the Hilbert space scales exponentially with the system size.

The language of complexity theory was first used to define integrability of quantum  many-body systems in \cite{caux2011remarks}. Let us briefly outline the construction introduced in \cite{caux2011remarks}.  A Hamiltonian of a quantum many-body system is defined on a Hilbert space which is a tensor product of a large number $N$ of elementary Hilbert spaces (e.g. Hilbert spaces of single spins $1/2$). This tensor product structure induces a natural notion of few-body (or few-site) operators, which act nontrivially  only in a fixed, independent on $N$ number of elementary Hilbert spaces.\footnote{A particular case of few-site operators are local operators, which act nontrivially on a fixed number of neighboring sites. QIMs  \eqref{I} of the PXP model  are local operators.}   A physical Hamiltonian itself is a sum of a polynomial (in $N$) number of few-body operators.  It was suggested in \cite{caux2011remarks} that a valid integral of motion should be representable as  a sum with at most polynomial in $N$ number of terms, each term being  a few-body operator \cite{caux2011remarks}.
This way one discards a large class of integrals of motion which bear little physical meaning in the many-body context, in particular, projection operators on individual eigenstates.

In line with the reasoning of Ref. \cite{caux2011remarks}, it is natural to require that functional dependencies relevant for defining integrability should also have at most polynomial  complexity.  This way one gets rid of functional dependencies with exponentially large number of terms like that in eq. \eqref{Neumann 1}.

\subsection{New definition of independence of QIMs\label{subsec}}

Now we are in a position to propose a definition of independence of QIMs of a many-body system which accounts for the aspects of the intuitive notion of such independence discussed above.

\medskip

\noindent {\it Definition 3.} Commuting integrals of motion $I_i,I_i,...,I_M$ of a many-body system defined over a tensor product of $N$ elementary Hilbert spaces are functionally independent unless there exists a function $F$ such that
\begin{itemize}
\item $F(I_1,I_2,...,I_M)=0$,
\item $F$ is nontrivial in the sense of {\it Definition 2},
\item  the complexity of $F(I_1,I_2,...,I_M)$, as defined in Ref. \cite{caux2011remarks}, is at most polynomial in $N$.
\end{itemize}

\medskip

\noindent This definition is the main result of the present paper. We refer to the notion of independence introduced in {\it Definition 3} as {\it independence in the strong sense} or, briefly, strong independence.

It should be stressed that the Definition 3 can not be applied  to a system with a fixed size $N$; it applies only to many-body systems in the limit of $N\rightarrow\infty$ (which can vaguely be referred to as thermodynamic limit). This feature comes bundled with the complexity requirement  \cite{caux2011remarks}. It has no counterpart in the definition of the classical integrability.

\subsection{A set of independent QIMs in the PXP model}

According to the Definition 3,  $N$ QIMs given by eq.  \eqref{I}  are not independent in the strong sense, due to the nontrivial functional relations \eqref{dependences}. Let us identify the size $\cal N$ of a maximal set of QIMs of the form    \eqref{I}  independent in the strong sense. To be specific, we assume that $N$ is a multiple of 3.

Eq. \eqref{dependences} implies that any three consecutive QIMs, $I_i,I_{i+1},I_{i+2}$, can not be independent in the strong sense. Thus  $\cal N$ can not exceed $2N/3$.

In fact, ${\cal N}=2N/3$, and one of the maximal sets of  QIMs independent in the strong sense includes all $I_i$ with  $i$ not divided by $3$, i.e.
$\{I_1,I_2,I_4,\dots,I_{N-1}\}$. To prove the strong independence of QIMs one notices that for any possible combination $\{\I_1,\I_2,\I_4,\dots,\I_{N-1}\}$ of eigenvalues of these QIMs there exists a common eigenstate  $\Psi_{\I_1,\I_2,\I_4,\dots,\I_{N-1}}$.
Assume there exists a functional dependence  $F(I_1,I_2,I_4,\dots,I_{N-1})=0$. Then
\begin{align}
0 & =F(I_1,I_2,I_4,\dots,I_{N-1})\Psi_{\I_1,\I_2,\I_4,\dots,\I_{N-1}}& \nonumber\\
&=F(\I_1,\I_2,\I_4,\dots,\I_{N-1})\Psi_{\I_1,\I_2,\I_4,\dots,\I_{N-1}} ,
\end{align}
and hence $F(\I_1,\I_2,\I_4,\dots,\I_{N-1})=0$. Since the latter equality is valid for an arbitrary combination of eigenvalues $\{\I_1,\I_2,\I_4,\dots,\I_{N-1}\}$, the functional dependence $F(I_1,I_2,I_4,\dots,I_{N-1})=0$ is trivial according to the {\it Definition~2}, and thus, according to the {\it Definition 3}, does not invalidate the strong functional independence of the set of QIMs under consideration.

Curiously, there exist sets of strongly independent QIMs of smaller sizes with the property that adding any single QIM of the form   \eqref{I} destroys the strong independence. One such set is the set of $N/2$  QIMs $\{I_{2i},\,i=1,2,\dots N/2\}$ (where $N$ is assumed to be even).


\bigskip

\section{Summary and concluding remarks\label{sec: summary}}

To summarise, we have introduced a new definition of functional independence of integrals of motion of quantum  many-body systems ({\it Definition 3}). We refer to this definition as independence in the strong sense, in contrast to the previously known independence in the weak sense  (see {\it Definition 1}). The new definition seems to be delicate enough to avoid various pitfalls discussed in the previous studies \cite{weigert1992,Sklyanin1992,gravel2002,Tempesta2004,caux2011remarks} and in the present paper. The definition imposes two novel requirements on a valid functional dependence of integrals of motion: The dependence should be nontrivial (as defined by {\it Definition 2}), and its complexity (as defined in Ref. \cite{caux2011remarks}) should scale at most polynomially  with the system size. Due to the later requirement the new definition in fact applies to a sequence of quantum many-body systems with growing sizes rather than to a system with a fixed size.

Armed with this definition, we have considered the  set \eqref{I} of integrals of motion of the PXP model \eqref{H} of $N$ spins~$1/2$.
This set contains $N$ binary integrals independent in the weak sense. We have found that the number of integrals of the form \eqref{I} independent in the strong sense is $2N/3$ (assuming $N$ is a multiple of 3). This leaves enough room for the nonintegrability, which indeed reveals itself by the Wigner-Dyson level statistics found in large subspaces of the Hilbert space of  the PXP model \cite{turner2018weak}.\footnote{Note, that, in principle, the PXP model can posses other few-body integrals of motion different from those in \eqref{I} (one such integral of motion is the Hamiltonian itself).}

It should be emphasized that independence in the weak sense, while insufficient to classify a many-body system as integrable,  remains a meaningful characteristics of a set of integrals of motion,  as is discussed in Sect. \ref{sec: common}.

Finally, we comment on the definition of integrability given in Ref. \cite{caux2011remarks}. This elaborated definition was crafted to accurately formalize intuition accumulated in studies of integrable quantum many-body systems. Yet, the PXP model \eqref{H} challenges this definition. According to the latter, the PXP model must be classified as integrable, since it possesses an extensive number of functionally independent local integrals of motion in involution. There are good reasons, however, not to regard the PXP model as integrable, as was discussed above.\footnote{Other nonintegrable models with an extensive number of independent local integrals of motion are also known \cite{hamazaki2016generalized,lan2017eigenstate,moriya2018}.} Thus the construction of Ref. \cite{caux2011remarks} needs to be amended in order to account for the PXP model and similar models.


%

\begin{acknowledgements}
Fruitful discussions with Boris V. Fine are acknowledged. The work was supported by the Russian Science Foundation under the grant No. 17-12-01587.
\end{acknowledgements}

\bibliographystyle{spmpsci}      



\providecommand{\noopsort}[1]{}\providecommand{\singleletter}[1]{#1}%



%
%

\end{document}